# Data science and Machine learning in the Clouds: A Perspective for the Future

Hrishav Bakul Barua[1]

*Robotics and Autonomous Systems Research Group*
*Cognitive Robotics & Vision*
*TCS Research*
*Kolkata, India*

**Abstract**

As we are fast approaching the beginning of a paradigm shift in the field of science, Data driven science (the so called fourth science paradigm) is going to be the driving force in research and innovation. From medicine to biodiversity and astronomy to geology, all these terms are somehow going to be affected by this paradigm shift. The huge amount of data to be processed under this new paradigm will be a major concern in the future and one will strongly require cloud based services in all the aspects of these computations (from storage to compute and other services). Another aspect will be energy consumption and performance of prediction jobs and tasks within such a scientific paradigm which will change the way one sees computation. Data science has heavily impacted or rather triggered the emergence of Machine Learning, Signal/Image/Video processing related algorithms, Artificial intelligence, Robotics, health informatics, geoinformatics, and many more such areas of interest. Hence, we envisage an era where Data science can deliver its promises with the help of the existing cloud based platforms and services with the addition of new services. In this article, we discuss about data driven science and Machine learning and how they are going to be linked through cloud based services in future. It also discusses the rise of paradigms like approximate computing, quantum computing and many more in recent times and their applicability in big data processing, data science, analytics, prediction and machine learning in the cloud environments.

## 1. Introduction

Big data is going to be the driving force for all the sciences in the earth [1]. Enormous data is being generated in the digital empire (in Zetta Bytes now) [2]. All the aspects of science specially medical research and astronomy heavily relies on the data generated from various sources. Data generation is not a problem but processing, storing, and retrieving the same is a huge problem in the current technological scenario. Even bigger is the problem of learning, analytics and prediction from such enormous amount of data. However, thanks to Cloud Computing [3] that the big data storage and analysis scenario is no longer a task as complex as it is supposes to be. But is Cloud Computing alone sufficient to address this shift of paradigm from Computational Science or less Data age to a exploratory science or Big data age [2]?

Classical data mining and analytics which primarily means single machine execution of data mining tasks, is the start of this revolution back in the 90's [4]. Single computer bound resources were capable to handle the amount of digital data and compute the same exactly. But as already stated by researchers that "*90% of world's data generated over last two years*" [5], one can never rely on classical techniques of data mining and analysis. Even Grid computing [6] based techniques fail big time. Cloud based services came as a boon to our digital world in 2000. All the essential elements of computation such as software, storage, security, infrastructure, server and platform can be availed as services in cloud today and that too in huge quantity as required. But again the question arise, is it enough to handle the upcoming Big data science era? No doubt that the cloud services mentioned have made data analytics and mining a trivial task with its vast expense of compute and storage resources although the management, concurrency, redundancy, load balancing, interoperability, and fault tolerance problems [7] are quite non-trivial to handle. But, what about energy and performance? Are we ready for the huge amount of energy required to process such huge data? The answer is NO! Unless we think out of the box and do something about it, we are going to land on a situation where we will have a shortage of energy and resources to devote in this new paradigm called data science.

The current research trends are concentrated towards a data-driven approach for almost everything. But to achieve the most out of data, we need to understand the relation between the various key influence areas of data science itself. Figure 1 depicts a search trend of some terms related to Data science and Cloud (retrieved from Google trends online application). Data science and Machine learning are on the rise as per people's interest over the time. There is a sudden hype of these terms since 2017.

---

[1]hbarua@acm.org

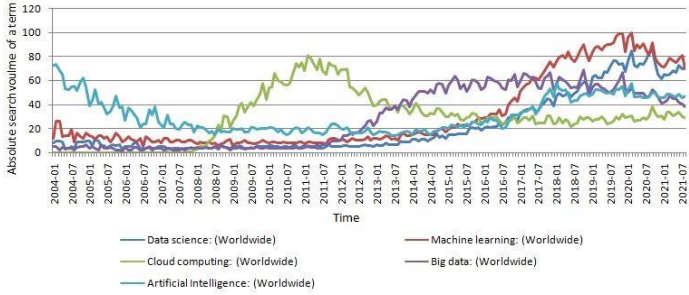

Figure 1: Search trends of Data science related terms as per Google trends application.

Cloud computing got a wave of interest between 2008 to 2013. But, now the interest is quite constant with time. Artificial intelligence, on the other hand, was popular from 2004 to 2008, then from 2008 to 2016 it was less searched, finally again from 2016 the interest has returned. Big data rose to prominence in 2012, but now the interest curve is in the downward direction. Data science returns about 2540 million records and Machine learning returns 2260 million in google search as of now. Big data returns 6720 million records which is quite a high amount. Cloud computing has 374 million records and Artificial intelligence has 907 million records. So, it seems that Big data has the highest number of records as per google search although the search interest is in the downward direction. Cloud computing and Artificial intelligence are less searched compared to others and record counts correspond to the search trend. Data science and Machine learning are well searched in recent times but the total records are less compared to Big data.

*Motivation and Research objective*. The main motivation behind the writing of this paper is to study and put forward the importance of Data science and Machine learning at large and use of Cloud based services, models, architectures, and technologies to facilitate their efficient use in terms of processing, analytics, prediction, inference and intelligence [8, 9]. We discuss various cloud based specialized services dedicated to the various aspects of Data science and Machine learning. We also put forward a collection of works in each of the category for the readers to perceive the topic effectively and get extended idea about the concerned areas. We also discuss a set of future possibilities in this respect. We try to attend to the following challenges and research questions in this review which find less mention in existing literatures.

- **RQ1.** What actually is Data science and how we have evolved towards this fourth science paradigm?

- **RQ2.** What are the current gaps between Data science and computation?

- **RQ3.** What are the Cloud services, platforms, frameworks and softwares for future Data science?

- **RQ4.** What are the emerging technologies (hardware or software) for efficient Data analytics and Machine learning in future Cloud systems?

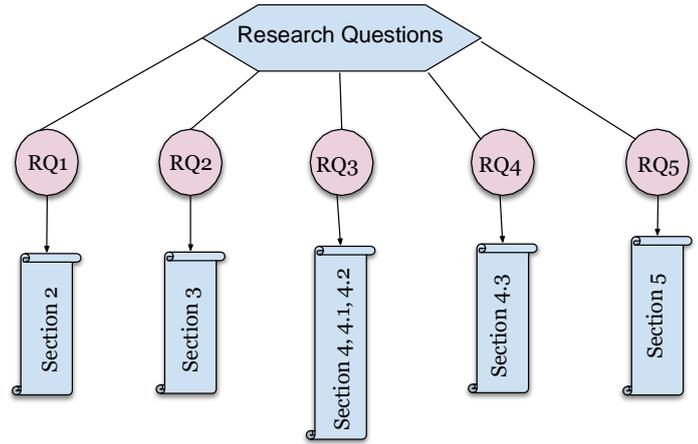

Figure 2: A depiction of the sections which answer the stated research questions.

- **RQ5.** What are the future directions and proposals in this area?

Figure 2 depicts the sections where each of the above mentioned RQs are answered.

*Organization of the paper*. The article is organized into the following sections. Section 2 gives a description and definitions of Data science and related fields. How it has evolved to the state we know today? The relation between various subfields of Data science is also mentioned. Section 3 gives a very brief background of Cloud computing and its services and its utility from the Data science perspective. Section 4 discusses the advances made on the field of Data science and Machine learning related computations atop the Clouds. Section 4.1 talks about the IaaS related entities and components for Data science and section 4.2 briefs about the PaaS related components for Data science. Some other cloud based specific and specialized services for Data science and related fields are discussed in section 4.3. The most important section of the article is section 5. In this section we propose service architectures for application specific jobs and tasks related to Data science in a holistic manner. The researchers, scientists, cloud engineers, data scientists and computer professionals can join hands to envision and realize such service specific architectures for better utility of cloud resources and components in the service of Data science. We conclude the article in section 6.

## 2. Data science and Machine learning: A Discussion

Data science [10, 11] is the so called science revolving around only one element called "Data". It also encompasses the methods, tools, and techniques required to handle, manage, retrieve, analyze and process data [9]. Now a days, this new kind of science has changed the world as we see today. Almost all the fields of science, engineering and even non-technical areas need data based information and intelligence to advance further. So, it is also called as the fourth science paradigm after experimental (first), theoretical (second) and computational (third) sciences before it. Figure 3 shows a rough timeline of evolution of science to the Data-driven science of today. But this field is highly interdisciplinary with players like statistics, applied mathematics, and domain specific knowledge contributing to its success along with computer science in the core. Big data is the driving force in Data science. The huge volume of digital data that we have at our disposal (see Section 1) is characterized by the 5 major Vs – Volume, Velocity, Variety, Veracity and Value.

Machine learning [12] is a sub-field of data science which primarily focuses on the fact that machines can learn how to recognize things in the surrounding through training from huge amount of data. The more the amount of data available for learning the merrier the outcome will be. This is quite similar to the fact that when a new baby is born he/she does not know anything about the mortal world. Gradually, by seeing, hearing and touching the elements of the surrounding environment he/she learns to recognize and act. The learning process is also facilitated by the parents' instructions and his/her own experiences through each passing year of his/her life. These are data-driven systems for achieving Artificial intelligence. The much complex tasks like Natural language processing, computer vision, scene understanding, 3D scene perception, signal processing, speech-to-text and vice versa systems are highly inspired by learning based paradigms.

Data science has ushered a whole new plathora of possibilities. Figure 4 shows some of the major sub-fields which are directly influenced by huge amount of data. Speaking in a generic terms, the two main sub-areas are – Data analysis [13, 14] and Artificial intelligence [15, 16]. The fields like Data analytics [17, 18] and Data mining [9] comes under Data analysis and Machine learning [12] and Deep learning [19] goes under Artificial intelligence. The major concern is the computational complexity of these methods and techniques in real world problems. The use of Neural networks and its variants like recurrent neural networks (RNN), convolution neural networks (CNN), deep neural networks (DNN), graph neural networks (GNN) and many more demand a huge amount of compute, storage and energy resources. It is not possible to establish such a setup in a standalone system or even if it is possible it is too costly to invest at all! Here we can chip in the idea of Cloud based resources and models to our advantage. The equations 1 and 2 give generic formulations for Big data and Data analysis in terms of other prominent elements of Data science (see [8] for details). Equation 3 shows the generic formulation of Artificial intelligence.

$$Data\ Analysis = Data\ Extraction + Data\ Cleaning + Data\ Transformation + Data\ Modeling + Data\ Mining + Data\ Analytics + Data\ Visualization + Inference + Conclusion \quad (1)$$

$$Big\ Data = Big\ Knowledge + Big\ Intelligence + Big\ Cognition \quad (2)$$

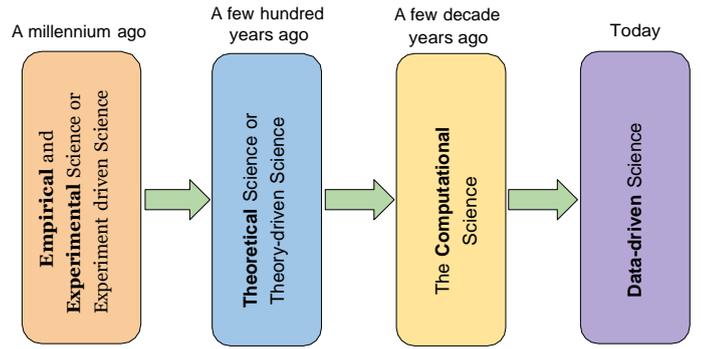

Figure 3: A timeline of evolution of Data Science, the fourth science paradigm.

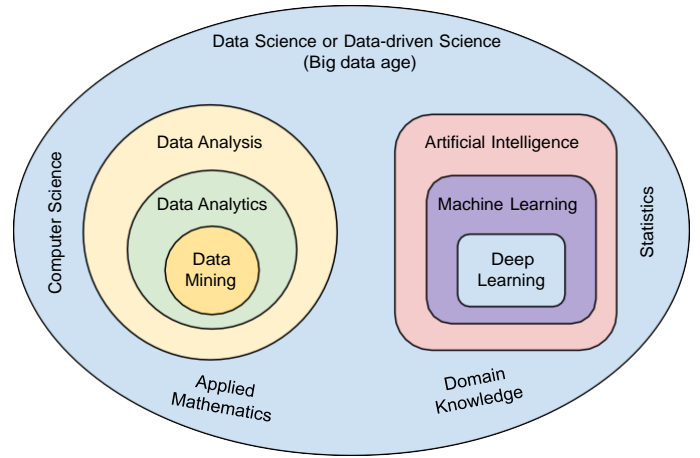

Figure 4: The holistic bird's eye view of Data science domain.

$$Artificial\ Intelligence = (Machine\ Learning + Deep\ Learning)\ on\ (Big\ Data) \quad (3)$$

Combining equations 1, 2 and 3 we get equation of Data science i.e. equation 4.

$$Data\ Science = Big\ Data + Data\ Analysis + Data\ Analytics + Artificial\ Intelligence \quad (4)$$

## 3. Cloud Computing services

Cloud computing [20, 21] is a novel idea of managing, utilizing and harnessing the wide range of hardware, software and firmware resources via network. It has become more important to consider Cloud computing paradigm seriously as computing has almost become the 5th utility after water, electricity, gas, and telephony/internet [20]. The main advantages of this paradigm is two fold. Firstly, it enables a user to choose from a vast sea of resources comprising of hardware (compute, storage, GPU, NPU etc.), Operating system, software, firmware, middle-ware, platforms and many more. Secondly, the model allows a per per use strategy for users, which means they can rent the resources for a fixed period of time, use them as per the needs and then give them back. So, users do not have to buy the resources on a permanent basis but just need to pay for the period of time they will be using the resources. The major service models of Cloud are – Software as a Service (SaaS) [22], Platform as a Service (PaaS) [23, 24], Infrastructure as a Service (IaaS) [25], and Storage as a Service (STaaS) [26, 27]. Figure 5 shows the major cloud service models available in the market. Table 1 presents the details of each of the major services mentioned above. STaaS is a service which comes within IaaS, so it is not mentioned as a separate entry in the table. But as it has come to much more prominence in the recent times due to emergence of Big data, hybrid storages, unstructured data management and storage, and object storage systems [28], so it is necessary to separate it out from IaaS. Apart from these there are many other job specific models which are being pursued in the recent times. Extensive discussion on the services and models related to Data science and the related fields are discussed as we go forward in the article.

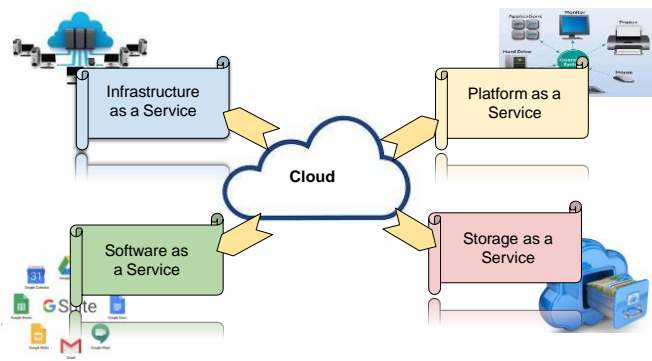

Figure 5: An overview of major cloud services available.

The big question is how to efficiently use these services to the advantage of Data scientific applications or how to redesign these service models to meet the exponential complexity growth in the 5 Vs of Big data. Can we create new concepts atop the existing ones without reinventing the wheel?

## 4. Cloud powered Data science and Machine learning

We have seen how Cloud computing can be leveraged to fulfill the computational needs in both processing and storage level [9, 8]. From the discussions on Data science and the related fields (see Section 2 and Figure 4) it is quite evident that to process jobs/tasks related to these technologies we need huge amount of computing power in terms of processors, Random Access Memory (RAM), Caches, Graphics Processing Units (GPUs) [29], Neural Processing Units (NPUs) [30], Quantum Processing Unit (QPU) [31] and many more such hardware units. Similarly, we need specialized storage systems for specialized tasks like Cloud storage [28], Object storage [32], File storage [33], Block storage [34], Archive storage [35] and so on. Now, utilizing the plathora of resources available to us itself is a challenge. Apart from that, to handle the huge volume of rapidly changing data (velocity) comprising of structured, semi-structure and un-structured data (variety), we require specialized frameworks and platforms which supports distributed computing, high degree parallelism and efficient storage management backbones. Table 2 presents the various fields of Data science (see Figure 4) with their definitions and some recent and major works in the Cloud environments.

### 4.1. Cloud Infrastructure resources for Data science

The various hardware resources included in IaaS (and STaaS) of the Cloud are used extensively by the Data science community. The specialized hardwares like GPUs [29], General Purpose GPUs (GPGPUs) [72], NPUs [30], QPUs [31], Cloud storages [28] and Approximate hardwares which can facilitate Approximate Computing (AC) [73] are utilized well in this regard. GPUs or Multi-GPUs are specifically designed for graphics related computations. But they can still be used for general computations to accelerate complex mathematical calculations which are often a large part of Data science and Machine learning. Moreover, computer vision related processing inherently requires multi-GPU systems for fast and efficient processing. GPGPUs are a specific class of GPUs which are dedicated towards general computing needs rather than vision and graphics processing. Also we have CPU-GPUs combining the usability of CPU and power of GPU for learning tasks. An NPU or neural processor is also known as an Artificial Intelligence (AI) accelerator. It is an Integrated Circuit (IC) that contains the necessary logic and controls to efficiently run Neural Networks (NNs) of any kind. Hence, it is well suited for Machine learning and Deep learning tasks like prediction and inference. Some other terms for NPU are Tensor processing unit (TPU), Intelligence processing unit (IPU), Vision processing unit (VPU) and Graph processing unit. QPU or Quantum chip is a processor chip that contains qubits. The qubits are the basic components of a quantum computing system. They remain interconnected in the NPU along with other control system components and supporting logic units. QPU is a powerful paradigm for executing complex mathematical models in a super fast manner. The concepts like Quantum Machine learning (QML) [74], Quantum Artificial Neural Networks (QANN) [75], and Quantum KNN (QKNN) [76] are already hot topics of discussion in the Data science community. Approximate Computing (AC) hardwares/tools are a class of special hardware units which are inexact or in-accurate in nature, i.e. they induce some tolerable error in the execution of the program or algorithm itself to reduce energy requirements and space/time complexity and bring greenness in computation [73]. Due to the inherent nature of Data science and Machine learning paradigms and NNs, approximating the results a bit does not harm the overall safety and outcome of the task. The approximate versions of NNs like – Approximate NN (AxNN) [77], Approximate LSTM (AxLSTM) [78], and Approximate SNN (AxSNN) [79] are already out in the market to explore. Table 3 lists the references for each of the discussed components in IaaS for Data science.

Table 1: Major Cloud computing based services

| Service | Definition | Elements | Examples |
|---|---|---|---|
| IaaS [25] | It is a cloud computing service which enables the delivery of servers (storage and compute), networking hardwares etc. as virtual machines to users. | Servers, Storage, Networking hardware, Virtual machines, etc. | Amazon Web services (AWS), Google Cloud/Compute Platform (GCP), Alibaba Cloud, Oracle Cloud Infrastructure, IBM Cloud, Microsoft Azure, Cisco Metapod, etc. |
| PaaS [23, 24] | It is a cloud computing service which enables the delivery of computing architectures, frameworks, development tools, analytics and other large scale services of various kind. | Operating systems, Data base management systems, Middle-ware, Firmware, etc. | AWS Elastic Beanstalk, Windows Azure, Google App Engine, OpenShift, Azure Cloud Services, etc. |
| SaaS [22] | It is the Cloud computing service which enables the delivery of web hosted applications centrally with proper licensing on subscription for end users. | Software, Applications, Tools, etc. | Microsoft Office 365, Dropbox, Google Applications (G suite), HubSpot, Shopify, Adobe Creative Cloud, and many more |

Table 2: Data science related fields, their definitions and references of works w.r.t. Cloud computing

| Field | Definition | References w.r.t. Cloud |
|---|---|---|
| Big data | It is the data which is too much huge, complex and rapidly changing in nature that conventional processing, mining and management systems are not sufficiently efficient to handle it | [9, 8, 36, 37, 38, 39, 40, 41, 42, 43] |
| Data analysis | The process of using traditional methods like classical statistics, applied mathematics and logic to extract meaningful information out of data | [44, 45, 46, 47, 48] |
| Data analytics | The process of gaining deep understanding and insights from huge amount of data using modern theories, tools, technologies, methods and systems (descriptive, predictive, discriminant, associative and prescriptive in nature) | [37, 8, 49, 50, 51, 52, 40, 41, 43, 53] |
| Data mining | The technique for finding meaningful information out of a vast expanse of data using traditional methods but for specific tasks | [9, 54, 55, 56, 57, 58] |
| Artificial intelligence | The intelligence in machines achieved primarily with audio-visual and touch perception (using microphone/cameras/other sensors) of the surroundings and cognition | [59, 60, 61, 42, 62] |
| Machine learning | The process of building computing models which learn themselves from huge amount of data and can perform perception/prediction/recognition tasks | [63, 64, 65, 66, 43, 67] |
| Deep learning | The process of building learning models that can closely imitate the working of human brain in perceiving the environment, decision-making and acting intelligently | [63, 68, 69, 70, 53, 71, 67] |

Table 3: References of Data science/Machine learning works based on Cloud infrastructures. For more explanations please see [73] and [28].

| Cloud infrastructure | Data science References |
|---|---|
| GPU, GPGPU, Multi-GPU, CPU-GPU | [80], [81], [82], [83], [84], [85], [86], [87], [88], [89], [90], [91], [92], [93], [94], [95], [96], [97], [84], [98], [90], [99], [100] |
| NPU | [101], [102], [103], [104] |
| QPU | [105], [75], [106], [107], [108], [105], [109], [74], [106], [110], [76], [111], [112], [113], [114], [107], [115], [108] |
| Approximate computing hardwares/tools | [101], [73], [116], [117], [118], [119], [120], [121], [122], [123], [124], [79], [125], [78], [77] |
| Cloud storage | [28], [126], [127], [128], [129], [130], [131], [132], [133], [134], [135], [136], [137], [138] |

*4.2. Cloud Platforms and Software resources for Data science*

The various resources included in PaaS and SaaS of the Cloud cannot be ignored when we are dealing with Data science and related fields. The frameworks, platforms, services and softwares for Data science and Machine learning in the Cloud are discussed here. A framework [9] is a way of execution of a certain program, algorithm or process. Data science related frameworks are specially designed in a way so that they can facilitate parallelism in the most granular level like Data parallelism, task parallelism and graph parallelism (for high volume data). The other important thing is scalability of tasks/jobs which are run atop such specialized frameworks. The frameworks must also support distributed computing for Machine learning tasks. The frameworks must be capable of handling variety of data types such as structured, unstructured and semi-structured. Also, considering velocity of Big data in today's world the frameworks must be able to handle streaming data as well. A platform [139] for Data science and Machine learning must contain the facilities for running Big data analytics and mining related jobs. Similarly, Deep learning paradigm support and other data intensive computational backbones should be an integral part of such a platform. The Neural network architectures like CNN, RNN, LSTM, etc. must be easily usable through such platforms. Such platforms must have a robust storage support at the back-end and attractive user's interface at the front-end. Cloud based Data science and Machine learning services [140, 139] are cost effective environments with a combination of compute, storage and other facilities dedicated for high performance computing (which generally comes under IaaS as discussed in Section 4.1). They often comprise of frameworks and platforms as a part of their service. So, basically speaking, algorithm runs atop a framework, a framework is a part of a platform, and finally a platform is facilitated by a cloud service for Data science. Figure 6 shows this concept clearly for the readers to perceive easily. The references to some of the important and relevant works in PaaS and SaaS pertaining to Data science can be seen in Table 4.

*4.3. Application specific Cloud services for Data science and Machine learning*

We have seen in the above discussions how Cloud based services like infrastructures, platforms, frameworks, storages and software are being harnessed to their full potential to meet the needs of Data science. But the question still remains, is this enough? Are we really at the optimum point of interest for a efficient, secure and robust Cloud based arrangement for Data scientific research, development and discovery?

With the increasing volume of Big data and its variety as well as velocity, processing is becoming costly and time consuming even after employing parallel, distributed or GPU computing facilities. Similarly the 2 other Vs – Veracity and Value also demands sufficient research and implementation attention from the concerned communities. What are the possibilities of defining and designing cloud based services specifically for Data science related endeavours like analytics, learning, prediction, mining, analysis, visualization, and many more of such kinds. Cloud based services and facilities are evolving and becoming more technology oriented day by day. It is really a surprise to see the emergence of a new kind of cloud service every now and then. Some of the recently proposed cloud models to facilitate specific applications over the cloud includes– Big data science as a Service [38], Analytics as a Service (AaaS) [177, 178], Cloud-Based Analytics-as-a-Service (CLAaaS) [179], Intrusion Detection System as a Service in Public Clouds (ID-SaaS) [180], Data as a Service (DaaS) [181], Big data as a service (BDaaS) [91], Health Informatics as a Service (HIaaS) [182], Quantum Computing as a Service (QCaaS) [183], Machine Learning as a Service (MLaaS) [184], Deep Learning as a Service (DLaaS) [185], Robotics-as-a-Service (RaaS) [186], Sensing-as-a-Service (SEaaS) [187], Drone-as-a-Service (DRaaS) [188], GPU-as-a-Service (GPUaaS) [189], Approximate Computing as a Service (AxCaaS) [73], Neural Processing as a Service (NPaaS) [104] and many more. All of these models are prepared on the basis of growing demand of the corresponding application areas and its impact on human life. Most of these services are directly or indirectly related to Data science and Machine learning applications so it is important to discuss some of them in a brief manner. The Table 5 summarises briefly the specialized cloud services related to Data science and the underlying basic services used in each case with the motivation behind its emergence. In the last row of the table, some hybrid services are proposed for Data science related activities and processes, which the researches can investigate in the future.

## 5. What is the future?

We have seen so far the importance of Cloud based resources and services for empowering the usage of Data science and Machine learning with explanations, citations of selected works and illustrations of concepts. But the future has much to unleash towards a more generic, useful and cost-effective cloud based architecture for Data science. Till now we see that there is a plathora of services in all the levels such as hardware/infrastructure, platform, framework/software and so on. All the services have their own advantages and features as we see in Table 5. So, to achieve a holistic service model which is dynamic in nature as well as can facilitate all the services pertaining to Data science is a future prospect for exploration. A similar Cloud architecture is envisioned in Figure 7. The figure is a bird's eye view which shows a layered architecture clearly distinguishing the different levels of services in the cloud – IaaS, PaaS and SaaS. Some major components and examples of of each of these services are included in the architecture which enables Data science related computations. Although it does not mention all the components due to space constraints but a holistic idea can be definitely drawn from it. The various components as stated in Tables 3 and 4 can be seen in the corresponding layers of the holistic architecture. The innermost layer, which corresponds to IaaS supports low level resources like GPUs, CPUs, NPUs, QPUs, Approximate hardwares (adders, multiplexers, storage, ALUs) and Cloud storages. All are interconnected via network elements and have storages linked with each of them in any cloud based distributed setup. A possible connection is shown among them (in Figure 5) although there can be many virtual setups for these hardwares in a logical mix and match fashion. The next layer supports PaaS for Data science with platforms like Amazon EC2, IBM Bluemix, Google Machine Learning Engine and so on. These platforms utilizes the infrastructure from IaaS efficiently to manage and process huge amount of data. These platforms also are highly capable of allowing frameworks and softwares in Data science and Machine learning to be incorporated on top of them. The next layer is the SaaS. This layer facilitates frameworks and softwares for Data science and Machine learning. These contains a wide range of examples like Hadoop, Spark and many more. The researchers and developers can utilize these SaaS components to develop and maintain algorithms with the support of PaaS based components. Finally, the outermost layer is a novel idea that the researchers can think for future generation cloud systems powering Data science. These services like – Data science as a Service, Machine Learning as a Service, Robotics as a Service and many more can be harnessed in the Clouds using dynamic combinations of components from each of the described layers of IaaS, PaaS and SaaS. Virtual environments and machines can be created using the IaaS components, then selected PaaS components can be utilized as per the service we are talking about and finally required SaaS components can be used atop PaaS to create a full fledged user/task specific service for Data science as listed in Table 5.

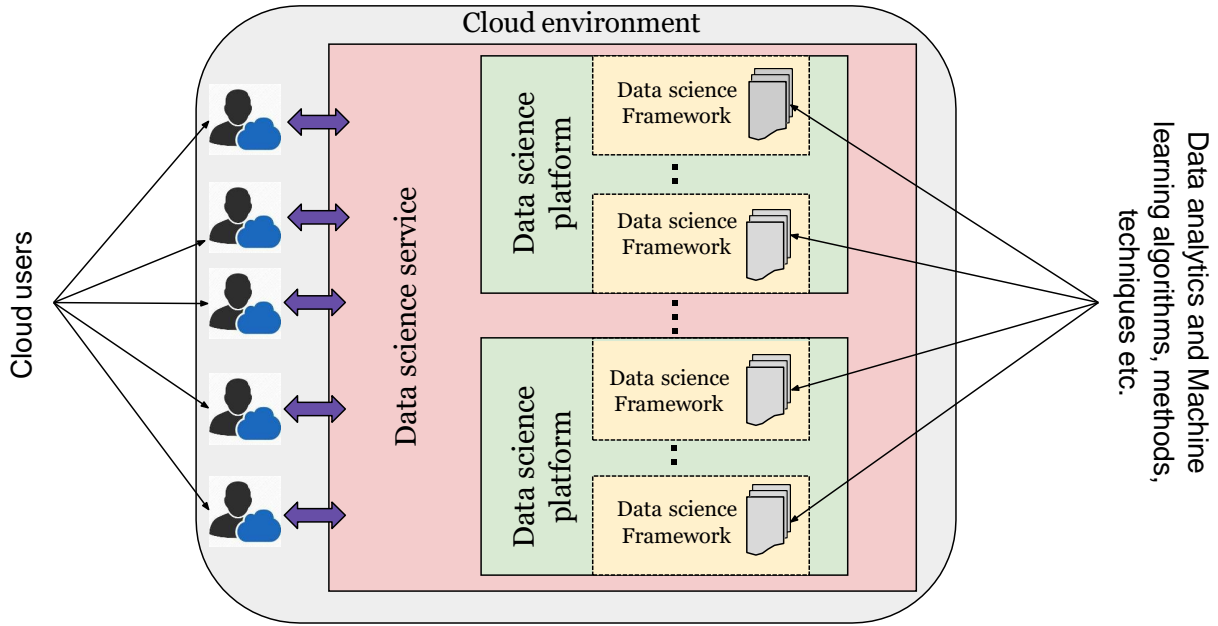

Figure 6: A generic setup showing the concept and relationship between Cloud based Data science services, platforms, frameworks/softwares and algorithms.

Table 4: References of Data science/Machine learning works based on Cloud platforms and frameworks. Please see [9] for more details.

| Cloud platform | Data science References |
| --- | --- |
| Platforms | [141, 142, 143, 144, 145, 139, 142, 146, 147, 148, 149, 150, 151, 152, 153, 154, 155, 100] |
| Services | [141, 144, 156, 148, 157, 150, 158, 159, 160, 161, 162] |
| Frameworks and Softwares | [163, 164, 165, 166, 167, 168, 118, 121, 169, 170, 171, 172, 173, 174, 175, 139, 146, 176, 100] |

Let us have a magnified look at a specific case for instance. Let us consider the cases in the last row of Table 5. We envision Machine Learning/Big Data services atop GPU, QPU, NPU and Approximate hardwares/tools. Figure 8 describes a future perspective on such a specialized service based cloud architecture. Each component of the architecture is numbered as you may see. (1) – The compute component of IaaS with CPUs, NPUs, GPUs, and QPUs (see Table 3). It also consists of Approximate hardwares. The various computing hardwares may be in a distributed virtual setup or inter-connected among themselves as per the requirement of the computation and data volume and velocity. Many combinations may be possible. (2) – The storage component which comes under STaaS of IaaS comprises of the various storage facilities of the cloud for storing any variety data such as structured, unstructured (images, videos and audios) and semi-structured. This is utilized as per the requirement of the Data science application in hand (see Table 3). (3) – This is the PaaS component of the architecture which enables run-time and developmental environments for Data science and Machine learning applications (see Table 4 and Figure 7). (4) – This component houses the SaaS in the form of frameworks, softwares and application specific to Big data and Machine learning (see Table 4 and Figure 7). (5) –

Table 5: Some Cloud Services on specific applications, their models and their motivations. Some short forms used in this table can be referred in the corresponding section 4.3

| Proposed cloud models | Cloud services in use | Other variants | Motivations behind | References |
|---|---|---|---|---|
| Big Data science as a Service | SaaS, PaaS, STaaS | CLAaaS BDaaS, HIaaS, DaaS, AaaS | Rising demand for high performance computing for big data management, mining, processing, visualization, analysis and analytics needs (considering the 3 major Vs) | [38, 177, 178, 179, 181, 91, 182, 190] |
| Machine Learning as a Service (MLaaS) | SaaS, PaaS, STaaS | DLaaS, IDSaaS | Meeting the exponentially rising demands for machine/deep learning models in most of the computing and AI related tasks such as NLP, deep computer vision, 3D scene generation in AR/VR technologies etc. | [184, 185, 180, 191, 192, 193] |
| Computing as a Service (CaaS) | PaaS, IaaS | QCaaS, AxCaaS, GPUaaS, NPaaS | Utilizing the recently rising computing paradigms in an efficient manner for optimizing energy requirements, space and time complexities | [189, 73, 183, 194, 195, 196, 104] |
| Robotics as a Service (RaaS) | IaaS, PaaS, SaaS | SEaaS, DRaaS | Integrating the robotics and other sensing/IoT hardware with web or cloud computing models for better and seamless usage | [187, 188, 186, 197, 198, 199] |
| Some other hybrid services | IaaS, PaaS, SaaS | ML-GPUaaS, ML-QCaaS BD-GPUaaS, BD-QCaaS, ML-AxCaaS, BD-AxCaaS, BD-NPaaS, ML-NPaaS | Combining the Big data (BD) analytics and Machine learning (ML) services with the power of GPUs, Approximate computing (AC), Neural Processing Units (NPUs) and Quantum computing (QC) paradigms | This is a prospective future scope of research and experimentation |

This is the most important component which demands active consideration and future vision for research. The main challenge is to mix and match among all the entities in each of the discussed components so far. How efficiently we can make suitable combinations and develop and design services which can handle Big data and Machine learning related work flows in a virtual machine setup of cloud. We can make unique combinations of *storage-compute-platform-software* as demanded by an application of Data science. In the process we can also design new platforms and services for specific research and academic communities. For example lets say we want to create a service for Machine learning jobs using GPUs or QPUs. So we make a combination of GPU-CPU in IaaS. In storage component we select unstructured data stores as ML generally uses this kind. In PaaS, we choose a particular ML based platform like Google Machine Learning Engine or may develop a new one to meet the new requirements. Finally in the SaaS stage, we utilize a suitable framework like spark or any newly developed software. At last all these entities in each of the components are stitched logically in a virtual machine setup and provided to the client. See equation 5 for the conceptual representation of one such example of an unique logical setup (see [8]). Figure 9 depicts such a specialized case.

$$(Acceleration + Efficiency + Greenness) \text{ in Data Science} = \\ Big\ Data\ in\ (\ GPU\ +\ QPU\ +\ NPU\ +\ AC\ +\ CPU \\ +\ Cloud\ Computing) \quad (5)$$

## 6. Conclusion

The article is an attempt to glance at the near future of Data science through the lens of Cloud. Data science has emerged as the fourth science paradigm which means it is going to impact all the other fields of science, technology and culture. It has become immensely important to bridge the gap between Data science and computation to allow efficient execution of complex Big data and Machine learning tasks. Cloud with its ocean of resources in various capacities is not a new concept but is it matured enough to be utilized for the huge amount of high velocity and variety data incurred by Data science and related fields. In this article we attempt to answer these questions for novice as well as expert researchers working in this domain. We attempt to envision some architectures for service specific models on Data science and Machine learning in a generic and holistic sense. The paper also surveys a collection of literature in each of the categories mentioned with an aim to facilitate the readers with a one stop index in this domain. The article shows the importance of hybrid service models in cloud for Data science and how it can bridge the gap between the computational and storage requirements of Big data and Machine learning applications.

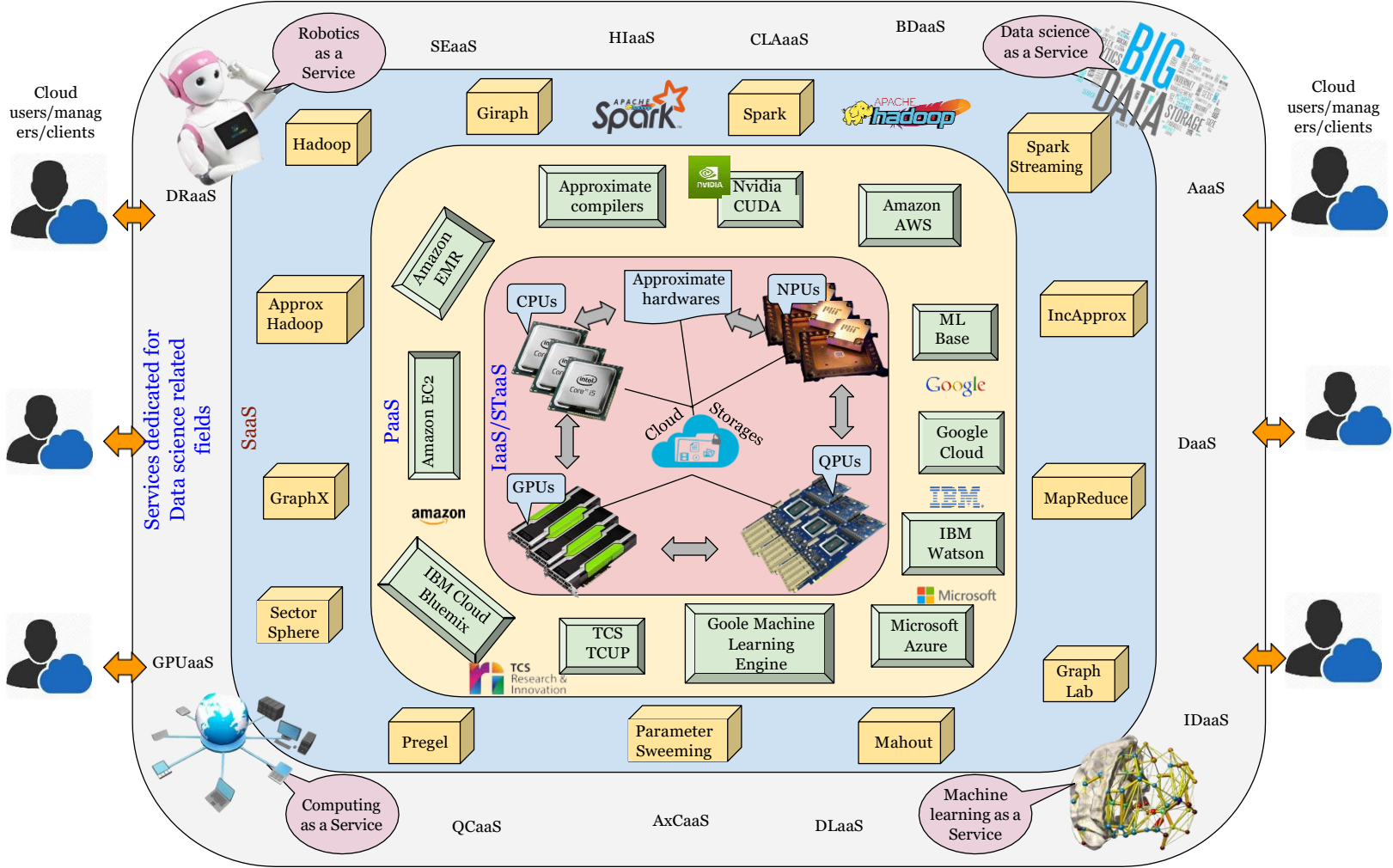

Figure 7: A generic and holistic proposed cloud architecture for Data science and Machine learning related platforms/services/infrastructures/frameworks. Some short forms are used in the figure for which the full forms can be found in the paper text.

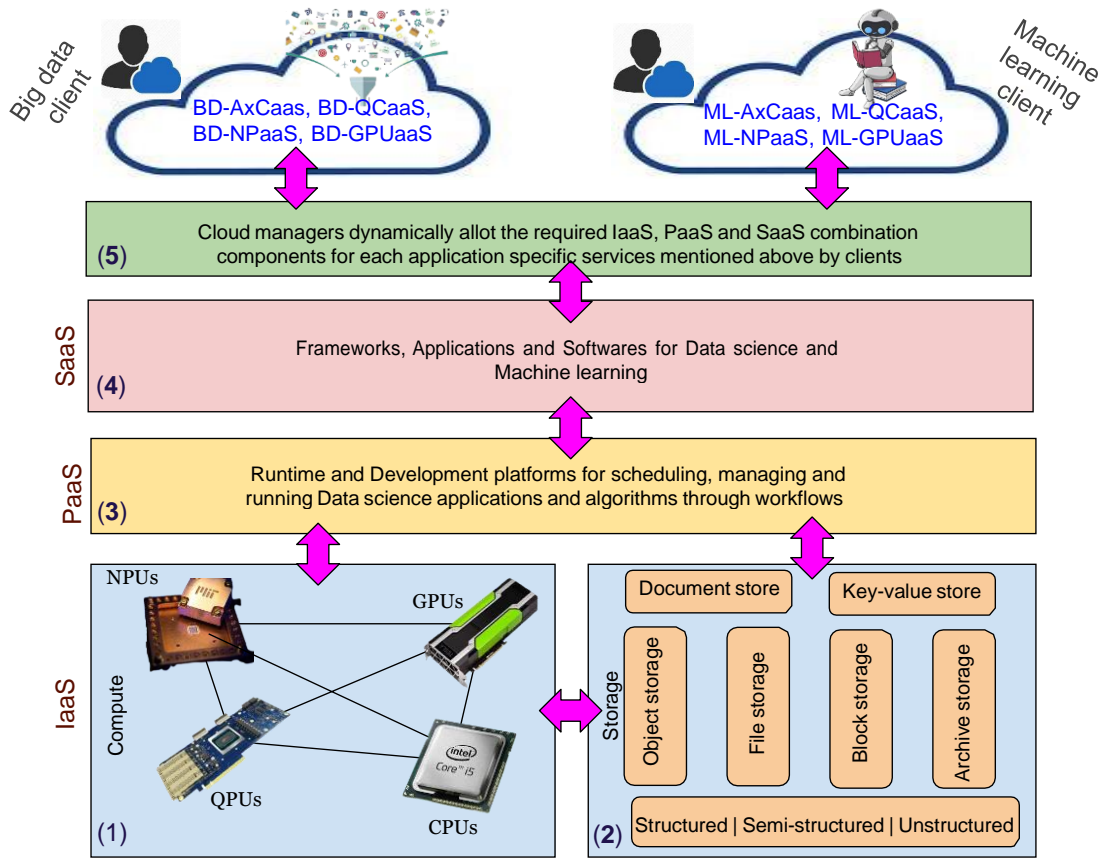

Figure 8: A specific proposed cloud architecture for Data science and Machine learning related platforms/services/infrastructures/frameworks. Some short forms are used in the figure for which the full forms can be found in the paper text.

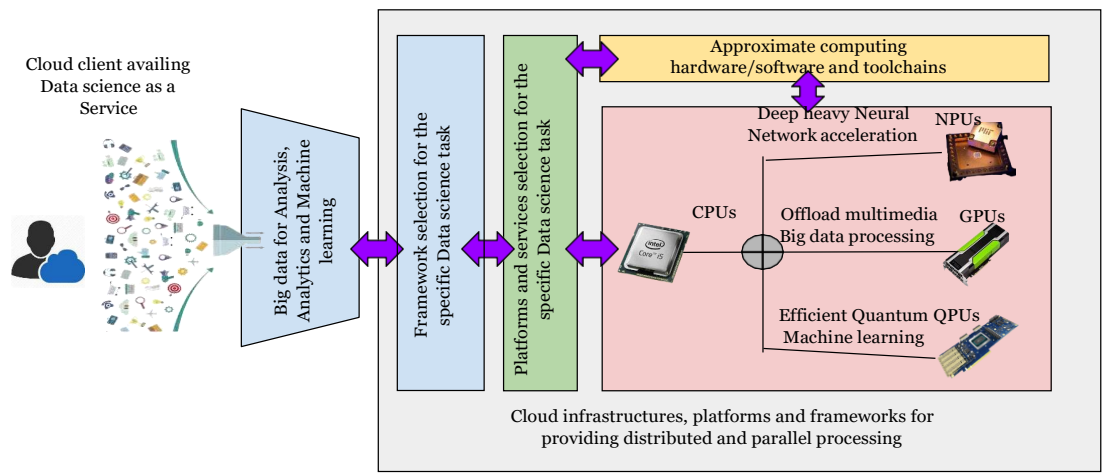

Figure 9: A very specific case of a Data science service combining various levels of Cloud based services as explained in equation 5